\documentclass[12pt, letterpaper]{article}
\usepackage{jheppubmodessay}
\notoc
\pdfoutput=1
\usepackage{hyperref}
\usepackage{psfrag}
\usepackage{array}
\usepackage{amssymb}
\usepackage{amsmath}
\usepackage{amsthm}
\usepackage{graphicx}
\usepackage{enumitem}
\usepackage{diagbox}
\usepackage[labelsep=quad]{subcaption}
\usepackage{cite}

\newcommand{\psine}{\Psi^{\text{ne}}} 

\def\psinesub[#1]{\psine_{#1}}

\newcommand{\cop}[1]{#1}
\newcommand{\al}{\cop{A}} 

\newcommand{\op}{{\cal O}} 

\def\Oright[#1]{\op_{#1}} 
\def\Oleft[#1]{\op_{L#1}} 
\def\anc{\cop{a}}

\def\an[#1]{\anc_{#1}} 
\def\aleft[#1]{\cop{a}_{L#1}} 
\def\arelr[#1]{\cop{a}^{\text{rel}}_{R#1}} 

\def\alen[#1]{\al_{\text{L}, #1}} 

\def\enrange[#1,#2]{{\cal R}_{#1}}
\def\dimrange[#1,#2]{{\cal D}_{#1}}
\def\hilb[#1]{{\cal H}_{#1}}

\def\projrange[#1, #2]{\cop{P}_{#1}}
\def\projh[#1]{\cop{P}_{\hilb[#1]}}

\def\proj[#1]{\cop{P}_{#1}}

\def\tarelr[#1]{\widetilde{a}^{\text{rel}}_{R#1}}
\def\coeff[#1]{\alpha_{#1}}

\def\pb[#1,#2]{\{#1, #2\}}
\def\deb[#1,#2]{[#1,#2]_{\text{D.B.}}}

\def\Or[#1]{{\text{O}}\left({#1}\right)}
\def\dotl[#1,#2]{\left\langle #1,\, #2 \right\rangle}
\def\dotlb[#1,#2]{\left\langle #1,\, #2 \right\rangle}
\def\dotlm[#1,#2]{\left[ #1,\, #2 \right]}
\def\dotp[#1,#2]{(\vect{#1} \cdot\vect{#2})}
\def\aff[#1,#2]{\hat{#1}(#2)}

\def\n4sym{{\cal N}=4 SYM}
\def\>{\rangle}
\def\<{\langle}
\def\weight[#1,#2,#3]{\{(#1),#2,#3\}}
\def\ads[#1]{$\text{AdS}_{#1}$}

\def\rtors[#1]{r_{*#1}}

\newcommand{\be}{\begin{equation}}
\newcommand{\ee}{\end{equation}}
\newcommand{\ba}{\begin{align}}
\newcommand{\ea}{\end{align}}
\newcommand{\bs}{\begin{split}}

\def\sess\end{split}
\newcommand{\vect}[1]{{\vec{#1}}}

\usepackage{setspace} 

\def\alcut[#1]{{\cal A}_{#1, \epsilon}}
\def\alseg[#1,#2]{{\cal B}_{#1, #2}}

\def\supcharge[#1]{\{#1\}}

\def\projsupeig[#1]{{\cal P}_{{\ell, m}}[{#1}]}
\def\transop[#1, #2]{T_{\{#1\}, \{#2\}}}
\def\supket[#1]{|\{#1\} \rangle}
\def\supbra[#1]{\langle \{#1\} | }

\title{How does information emerge from a black hole?}
\author{Suvrat Raju}
\affiliation{International Centre for Theoretical Sciences, Tata Institute of Fundamental Research, Shivakote, Bengaluru 560089, India.}
\emailAdd{suvrat@icts.res.in}
\date{}
\abstract{ An examination of the constraints of quantum gravity leads to a clear physical picture for how information about the initial state is transferred to the Hawking radiation that emerges from a black hole.

\vskip 1in
}

\begin{document}
\maketitle
\section{Introduction}
It is now generally accepted that black-hole evaporation preserves unitarity. Evidence from the AdS/CFT conjecture \cite{Maldacena:1997re,Witten:1998qj,Gubser:1998bc} shows that black-hole evaporation in asymptotically anti-de Sitter space (AdS) can be modelled using a dual unitary process.  The key information-theoretic aspect of this duality can be generalized to asymptotically flat space \cite{Laddha:2020kvp} and asymptotically de Sitter space \cite{Chakraborty:2023los} through the principle of ``holography of information'' \cite{Laddha:2020kvp,Chowdhury:2020hse,Raju:2020smc,Chowdhury:2021nxw,Raju:2021lwh,Chakravarty:2023cll, Chakraborty:2023los, deMelloKoch:2022sul,Chowdhury:2022wcv,Gaddam:2024mqm}. These results, which build on \cite{Marolf:2008mf,Marolf:2013iba,Banerjee:2016mhh,Jacobson:2012gh,Jacobson:2019gnm},  imply that, subject to reasonable assumptions about the global state and the UV-completion of low-energy quantum gravity,  information in a bounded subregion of a Cauchy slice can be accessed by observables in its complement.  Since this principle applies even when the subregion is part of a black hole,  an external observer can always verify the purity of the state. This provides a resolution of the information paradox \cite{Hawking:1974sw,Hawking:1976ra} for evaporating black holes as explained in \cite{Raju:2020smc,Raju:2021lwh}. 

Additionally, in separate calculations involving AdS black holes coupled to nongravitational baths \cite{Penington:2019npb,Almheiri:2019qdq,Almheiri:2020cfm,Almheiri:2019psf}, the entanglement entropy of the bath has been shown to obey a Page curve \cite{Page:1993df}.

Nevertheless, the literature lacks clarity on how information from the infalling matter emerges in the final Hawking radiation. This has led to speculations that the process is mediated by exotic  wormholes, or modifications of the semi-classical black-hole geometry \cite{Mathur:2005zp}.  The purpose of this chapter is to explain that the transfer of information can be  understood through an analysis of the constraints of quantum gravity. This mechanism does not require us to invoke new physical phenomena.

\section{The classical Gauss law}
Let us start with an elementary but instructive discussion. Here, and in most of  what follows, we will restrict ourselves to asymptotically flat space. Consider a star  that collapses into a black hole and then evaporates. One piece of information  present in the state is the total energy (or mass) of the star. Nevertheless, there has never been any dispute that an observer outside the black hole retains access to  this information even after the matter that constitutes the star has disappeared behind the horizon. This is because the energy can be obtained by integrating a component of the metric at infinity using the ADM formula \cite{Arnowitt:1962hi},
\be
\label{admh}
H = {1 \over 16 \pi G} \int_{S_{\infty}} \left(\partial_i g^{i}_{j} - \partial_j g_{i i} \right)  d^2 \Omega^j.
\ee
This is a manifestation of the Gauss law.

The precise expression for the ADM Hamiltonian involves observables at infinity.  Since this chapter aims at providing physical pictures,  we will somewhat loosely state that the metric ``outside'' the black hole has information about its energy. 

Even in this context, one might ask: how does information about the energy enter the final Hawking radiation? The matter that sources this energy is causally disconnected from the region outside the black hole, where the radiation forms.  But information about the energy does not ``jump'' from the matter to the exterior.  Rather, one must recall that the matter came from outside the black hole in the past. As it collapsed, it modified the exterior metric, and a part of the  metric continues to retain information about the total energy. Since the exterior metric carries information about the energy at leading order, it is not surprising that Hawking radiation from the black hole also carries information about the energy at leading order.

Moreover,  information about the energy is not ``cloned'' in the metric at infinity. Rather the gravitational constraints instruct us to {\em identify} the energy in the bulk with the observable at infinity: the seemingly different ways of measuring the energy correspond to the same operator in the physical Hilbert space.

Of course, in the classical theory, observations outside the black hole are insufficient to fix the state of the interior. Classically, it is easy to construct two distinct states, for which not just the energy but all multipole moments coincide.  The classical no-hair theorem suggests that as long as we treat the gravitational theory classically --- as is done in textbook analyses of Hawking radiation that utilize quantum field theory in a fixed spacetime background --- the Hawking radiation only carries information about the energy of the black hole and about other conserved charges that affect the classical metric.

\section{Holography of information}
When quantum effects are included, it turns out that observations outside the black hole {\em are} sufficient to determine its complete state. The precise results proved in \cite{Laddha:2020kvp,Chakraborty:2023los} pertain to asymptotic algebras. 
But their physical interpretation  is that an observer outside the black hole can completely determine the state of the interior.

These results are essentially based on the Gauss law and the uncertainty principle.  The underlying physics can be understood using the following simple argument.   Consider a nice slice that passes through the black hole interior and extends out to spatial infinity. Say that there exists an operator in the theory that can be used to alter some observables in the part of the slice inside the black hole while commuting with all operators outside. Such an operator must commute with the Hamiltonian, which is observable outside in a gravitational theory.  However, the uncertainty principle tells us that such an operator, which has ``zero'' energy,  must be delocalized. But then it must fail to commute with some {\em other} operator in the exterior. Therefore, such an operator cannot exist at all.

This is related to the usual argument that there are no gauge-invariant local operators in a theory of gravity. However, the physical consequences of this statement are not widely appreciated. For low-energy excitations about the vacuum, observers outside a bounded region can follow  a well-defined perturbative protocol that completely identifies the state in the region \cite{Chowdhury:2020hse}. Moreover, by adding weak assumptions about the UV-complete theory, it is possible to obtain a nonperturbative formulation of the principle of holography of information for asymptotic algebras \cite{Laddha:2020kvp}. This nonperturbative result can be applied to black holes. We describe these regimes in turn below.

\subsection{Perturbative example \label{subsecpert}}
\begin{figure}[!h]
\begin{center}
\includegraphics[height=0.3\textheight]{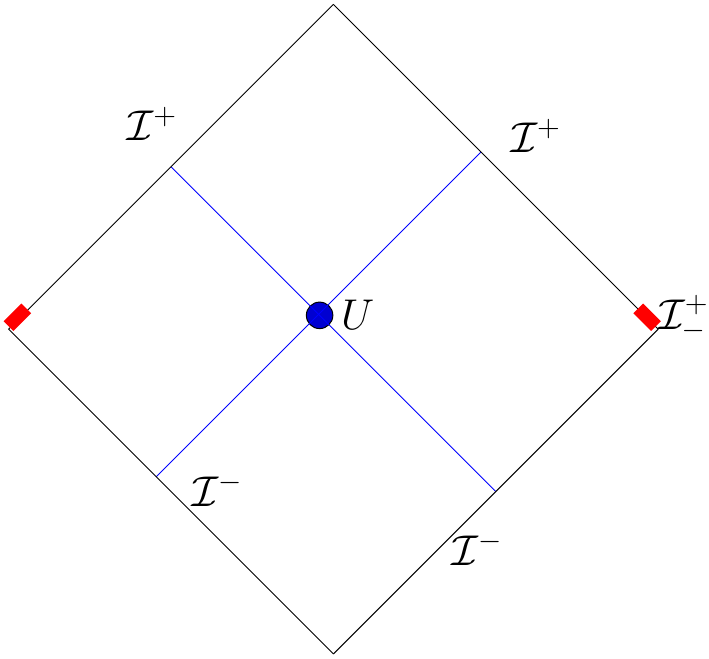}
\caption{\em The blue excitation can be written as the action of a ``local'' unitary on the vacuum but it must originate from ${\cal I}^{-}$ in a gravitational theory. This process leaves behind traces at infinity that are sufficient for an observer near ${\cal I}^{+}_{-}$ (red region) to determine  the profile of the excitation.  \label{pertdetection}}
\end{center}
\end{figure}

A simple perturbative example, depicted in Figure \ref{pertdetection}, illustrates this phenomenon about empty space.  Let $\phi$ be a massless scalar field coupled to gravity. Consider the following unitary operator that acts near the ``middle'' of the spacetime to create an excited state from the vacuum,
\be
|f \rangle = e^{-i \lambda \int f(r) \phi(u=0,r, \Omega) d^2 \Omega d r} | 0 \rangle,
\ee
where $u = t - r$ is the retarded time. 
The information in this state is characterized by the smearing function $f(r)$ that we take to have compact support in the range $r \in (0,r_0)$ for some $r_0$.   We also take $f(r)$ to be spherically symmetric for this example; generalizing to an angle-dependent smearing is straightforward. 

Although this unitary appears to be localized, it can be detected by an observer near infinity. Consider an observer who can make observations near the past boundary of future null infinity (${\cal I}^{+}_{-}$): the region obtained by first taking $r \rightarrow \infty$ and then taking $u$ to be negative and large in magnitude. Such an observer can compute the correlator of the Hamiltonian, which is available at infinity in a theory of gravity,  and another insertion of the field near ${\cal I}^{+}_{-}$.
\be
\lim_{r' \rightarrow \infty} r' {\partial \over \partial \lambda} \langle f | H \phi(u', r',  \Omega') | f \rangle \big|_{\lambda=0} = {1 \over 2 \pi} \int_0^{r_0} {f(r) \over u' (u' - 2 r)}  d r.
\ee
It is easy to see that a knowledge of the right hand side for the range $u' \in (-\infty, -{1 \over \epsilon})$, with any finite $\epsilon$, is sufficient to completely reconstruct $f(r)$. For instance, series expanding the denominator in ${r \over u'}$  yields all moments of $f(r)$ as coefficients of different inverse powers of $u'$,
\be
{1 \over 2 \pi} \int_0^{r_0} {f(r) \over u' (u' - 2 r)}  d r = \sum_{n=0}^{\infty} {2^n \over 2 \pi (u')^{n+2}} \int_0^{r_0} r^n f(r) d r,
\ee
and these moments determine $f(r)$ to any required precision.

Therefore, we see that an observer at infinity has all information about the state. But the mechanism by means of which this happens is not mysterious.  In a theory of quantum gravity, there is no such thing as a local source.  Therefore, the state $|f \rangle$ must be thought of as originating on ${\cal I}^{-}$. As this excitation enters the bulk, it modifies the correlators at infinity in a manner that is consistent with the gravitational constraints. These modified correlators retain information about the state even after the excitation has seemingly fallen in, and before it emerges out to ${\cal I}^{+}$. 

In ordinary nongravitational theories (including nongravitational gauge theories), it is possible to prepare a state by sending in information into a bounded region, while erasing all traces of this information outside the region.  In a theory of gravity, the principle of holography of information tells us that this is impossible.

The example can be embellished to display some additional instructive features. Consider the case where the theory has a second field $\widetilde{\phi}$ that couples to gravity in precisely the same manner as $\phi$. We can take $\phi$ and $\widetilde{\phi}$ to be related by a global symmetry and set the $\phi-\widetilde{\phi}$ two-point function to vanish.  (Theories of quantum gravity might not display such global symmetries but this issue is irrelevant for the present discussion.)

Now consider the state
\be
|\widetilde{f} \rangle = e^{-i \lambda \int f(r) \widetilde{\phi}(u=0,r, \Omega) d^2 \Omega d r} | 0 \rangle,
\ee
which is similar to the state $|f \rangle$ defined above, but contains an excitation of $\widetilde{\phi}$ rather than $\phi$. 

It might naively appear that the observer at infinity cannot distinguish $|f \rangle$ and $|\widetilde{f} \rangle$. However, one finds that
\be
\lim_{r' \rightarrow \infty} r' {\partial \over \partial \lambda} \langle \widetilde{f} | H \widetilde{\phi}(u', r',  \Omega') | \widetilde{f} \rangle \big|_{\lambda=0} = {1 \over 2 \pi} \int_0^{r_0} {f(r) \over u' (u' - 2 r)}  d r.
\ee
This is similar to the result above except that it is $\widetilde{\phi}$ that has been inserted in the state $|\widetilde{f} \rangle$. 
Moreover,
\be
\lim_{r' \rightarrow \infty} r' {\partial \over \partial \lambda} \langle \widetilde{f} | H \phi(u', r',  \Omega') | \widetilde{f} \rangle \big|_{\lambda=0} = \lim_{r' \rightarrow \infty} r' {\partial \over \partial \lambda} \langle f | H \widetilde{\phi}(u', r',  \Omega') | f \rangle \big|_{\lambda=0} = 0
\ee

Therefore, an observer at infinity can distinguish the two states by determining whether the two point function of $H$ and $\phi$ is nonzero or whether the two point function of $H$ and $\widetilde{\phi}$ is nonzero. 

\subsection{Nonperturbative results \label{subsecnonpert}}
The perturbative examples provided above should not be thought of as a proof of the holography of information. A nonperturbative result for asymptotic algebras is proved in \cite{Laddha:2020kvp}. (See \cite{Raju:2020smc} for a simplified discussion.) In the case of asymptotically flat space, the result is that any observable on ${\cal I}^{+}$ can be approximated arbitrarily well by an observable near its past boundary, ${\cal I}^{+}_{-}$. Correspondingly, any observable on ${\cal I}^{-}$ can be approximated arbitrarily well by an observable near its future boundary ${\cal I}^{-}_{+}$. 

Since we do not know the full UV-complete theory of quantum gravity in asymptotically flat space, this result requires some assumptions about that theory. The assumptions required are as follows.
\begin{enumerate}
\item
The formalism of quantum field theory is applicable in the asymptotic region and the asymptotic algebra continues to make sense in the full UV-complete theory.
\item
The Hamiltonian is bounded below.
\item
The algebra at infinity (more precisely, the algebra at ${\cal I}^{+}_{-}$ or ${\cal I}^{-}_{+}$) contains projectors onto the vacua of the theory.
\end{enumerate}
The first assumption is reasonable if one adopts the perspective that, even in quantum gravity, the asymptotic geometry should be held fixed. Therefore, while quantum field theory might not provide a useful description in the bulk, one still expects it to provide a useful description asymptotically. The second assumption is difficult to prove but is reasonably expected in any quantum-mechanical theory. (For convenience, we set the lower-bound of energy to be zero.)  The third assumption is related to the Gauss law. In the low-energy theory, the energy can be measured at infinity and the projector onto the vacuum is simply the probability that one obtains zero in such a measurement. The vacuum might be degenerate, as in four-dimensional flat space but, in that case, the supertranslation charges that identify the different vacua are also elements of the algebra at infinity in the low-energy theory. The third assumption states that this property extends to the UV-complete theory.

\subsection{Intermediate regime}
The principle of holography of information is important in some situations and not in others.  An analogy with our understanding of unitarity in ordinary quantum field theories might help to illustrate this point. In the canonical formalism, it is possible to formally prove that the S-matrix is unitary. Unitarity can be verified through explicit calculations when one studies the scattering of simple excitations of the vacuum.  However, in the presence of a heavy nonperturbative background with  $e^{S}$ microstates,  effective field theory appears to be dissipative unless one keeps track of effects of size $\Or[e^{-S}]$.  

In precisely the same spirit, discussions of holography in gravity should be separated into three regimes. The nonperturbative regime alluded to in section \ref{subsecnonpert} provides the correct setting to ask fine-grained questions, such as those about the unitarity of black-hole evaporation. The second regime is where one studies simple states in perturbation theory, without requiring any assumptions about the UV-complete theory; this is described in section \ref{subsecpert}.  A third, intermediate, regime is where one introduces a heavy background with $S$ microstates and coarse-grains the algebra of operators \cite{Bahiru:2022oas,Jensen:2024dnl} by discarding observables suppressed by $S$. In this regime, holography is obscured and gravity behaves like other quantum field theories.  This is consistent with our mundane experience.  

The constructions of  \cite{Bahiru:2022oas,Jensen:2024dnl} that build on \cite{Papadodimas:2015xma,Papadodimas:2013jku} effectively ``dress'' bulk excitations to the background state so as to obtain local operators that perturbatively commute with operators at infinity. This extends the standard procedure that is used to obtain quasilocal operators in heavy classical backgrounds \cite{DeWitt:1967yk}. 

However, a study of the third regime should not lead one to the erroneous conclusion that holography of information is a property of ``special states'' \cite{Bahiru:2022oas}. This would be analogous to studying coarse-grained models that are dissipative and using them to conclude that unitarity is only a property of ``special states''. 

\section{A physical picture for information transfer}

The effect described above gives us a clear picture of how information emerges from the black hole. At leading order, one-point functions of the metric and other fields outside the black hole have information about the mass, charge and angular momentum. So the Hawking radiation formed in this background carries this information  at leading order. Quantum effects in this system are naturally controlled by the parameter $G E^{2} \propto {1 \over S}$, where $E$ is the energy of a Hawking quanta and $S$ is the entropy of the black hole. Additional information about the black hole is encoded in ${1 \over S}$-suppressed observables outside the black hole and this affects subleading features of the Hawking radiation.  For instance, the Hawking radiation from a black hole with ``canonical spread'' in energy differs, at $\Or[{1 \over S}]$,  from the radiation that emanates from a black hole with ``microcanonical spread'' in energy.

The purity of the state can be verified at all times outside the black hole,  provided that one keeps track of effects of size $\Or[e^{-S}]$.
This is perfectly consistent with the usual statistical arguments which suggest that it is necessary to probe the final radiation to nonperturbative order to verify that its form is consistent with unitarity  \cite{Raju:2020smc}.

\begin{figure}[!h]
\begin{center}
\includegraphics[height=0.4\textheight]{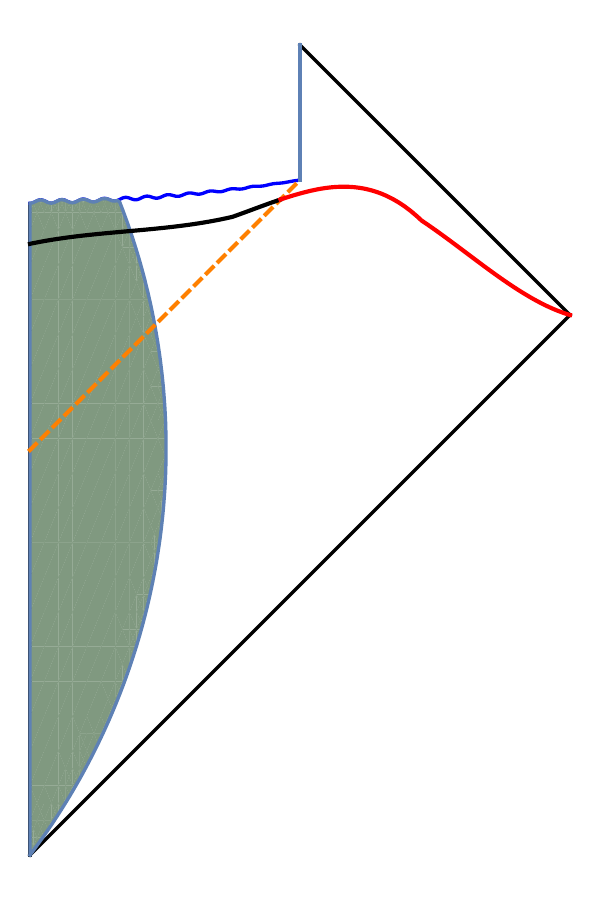}
\caption{\em In the classical theory, observables on the red-part of the Cauchy slice (outside the horizon)  have information about the total energy. In the quantum theory, observables in this region suffice to completely identify the state of the black hole.  In both cases, the mechanism is that the process of black-hole formation modifies observables in the black-hole exterior, and the gravitational constraints force them to retain information about the infalling matter \label{exteriorobserver}}
\end{center}
\end{figure}
To summarize: the mechanism by which information is transferred to the Hawking radiation is that the constraints of quantum gravity force it to always remain available outside the black-hole horizon.  Because fine-grained details of fluctuations of the metric outside the horizon and the correlations of these fluctuations with other observables are sensitive to the state of the infalling matter, it is natural that the Hawking radiation contains this information after the black hole has evaporated.  Figure \ref{exteriorobserver} summarizes the main point of our discussion.

\section{Comparison with other pictures}
The mechanism described in this chapter has some features in common with other proposed resolutions of the information paradox. In this section, we present a brief comparison.

\paragraph{\bf Fuzzballs.}
The flavour of the ``fuzzball resolution'' is broadly similar to ours since it also suggests that the wavefunctional outside the horizon provides a complete description of the state. However, the framework presented above does not require us to modify the metric at the semi-classical level. Instead, we expect that the expectation value of the metric will coincide up to exponential accuracy with the usual black-hole metric.   It is only small quantum features of observables  outside the black hole that retain information about the initial state. The fuzzball proposal was motivated by a search for loopholes in the classical no-hair theorem.  But the no-hair theorem fails as soon as gravity is treated quantum mechanically and one need not seek any modifications at the classical level. 

\paragraph{\bf Soft hair.}
The quantum observables outside the black hole that suffice to distinctively characterize its state can be thought of as ``quantum-mechanical hair.'' This hair arises automatically when one correctly imposes the constraints of gravity in quantum
mechanics.   In \cite{Hawking:2016msc}, it was proposed that soft hair on black holes might help preserve information about the initial state. But the principle of holography of information  does not rely on infrared effects; in fact, the presence of soft hair was a complication that had to be surmounted in \cite{Laddha:2020kvp}. For this reason, our discussion generalizes nicely to AdS where the cosmological constant regulates infrared effects.

\paragraph{\bf Measurements of energy}
An old proposal to resolve the information paradox \cite{Balasubramanian:2006iw} suggests that if the energy eigenstates are discrete, and energy can be measured at infinity with exponential precision then this can be used to obtain information about the microstate. 

While this proposal was prescient, correlators of $H$, by themselves, are insufficient to determine the state. For instance, correlators of $H$ cannot even distinguish between the pure state,
\be
|\Psi \rangle = \sum_{E} c_{E} | E \rangle,
\ee
written as a linear combination of energy eigenstates, and the mixed density matrix,
\be
\rho = \sum_{E} |c_{E}|^2 | E \rangle \langle E |. 
\ee
As even the perturbative example shows, it is necessary for the distant observer to study observables that do not commute with $H$ in order to identify the state.


 

\paragraph{\bf Quantum hair.} The authors of \cite{Calmet:2021cip,Calmet:2021stu} arrived at the same conclusion as earlier work \cite{Raju:2020smc,Raju:2021lwh} that is summarized here: information about the black-hole interior is always present in the exterior. However, the techniques used in \cite{Calmet:2021cip,Calmet:2021stu} are different from those used to prove the holography of information and, as such, they are insufficient to establish this conclusion. 

The authors of \cite{Calmet:2021stu} use an ``effective action'' for gravity which suggests that subleading terms in gravitational field at infinity  are sensitive to the details of the density distribution of a spherical source. Regardless of the validity of this effective action, presumably, the metric obtained by this procedure should be thought of as the quantum-corrected expectation value of the metric operator. However, in general, one requires quantum correlations of the metric and other observables at infinity to obtain complete information about the state. It is insufficient to examine one-point functions of the metric as is already clear from the perturbative examples presented above.  

Second, the matter fields in \cite{Calmet:2021stu} are also not treated quantum mechanically. In \cite{Calmet:2021stu}, it is suggested that the matter be placed in an ``energy eigenstate''. However, the uncertainty principle implies that the wavefunctions of an energy eigenstate are always delocalized, even in the absence of gravity. So energy eigenstates should not be used as models of localized excitations. 

Third, the constraints of gravity correlate the metric to the state of the matter field. Therefore, in the gravitational theory, a state corresponding to a superposition of matter excitations should be described as a sum of ``dressed states'' where, in each term, the state of the constrained components of the metric is related to the state of the dynamical components of the metric and other fields. But the starting point of \cite{Calmet:2021cip} is a simple superposition of geometries, which is incorrect. 

For these reasons, we direct the reader to \cite{Chowdhury:2021nxw}, where the metric and matter fields are treated quantum mechanically and the constraints are solved to obtain the full wavefunctional. This shows that observables at infinity (see Eqn. 6.6 in \cite{Chowdhury:2021nxw} for a precise list of observables) uniquely fix the bulk state.  However, we caution the reader that while the techniques of \cite{Chowdhury:2021nxw} are sufficient and instructive at the perturbative level,  they cannot easily be extended to the nonperturbative setting, which is relevant for black holes. In that setting, the arguments of \cite{Laddha:2020kvp} are required.

\paragraph{\bf Page curve.}
Page \cite{Page:1993wv} argued that information is initially lost in the black hole and then starts to emerge gradually after about ``half'' the black hole has evaporated. In contrast, the result utilized here is that the observer outside can always determine that the state is pure.  This result is not in contradiction with recent derivations of the Page curve \cite{Penington:2019npb,Almheiri:2019qdq,Almheiri:2020cfm,Almheiri:2019psf}. An essential feature \cite{Geng:2020qvw,Geng:2020fxl,Geng:2021hlu} of these Page curves is that they are obtained by adding a nongravitational bath to an AdS black hole. They do not  measure how information emerges from the black hole but merely how it flows from one part of the nongravitational bath to the other.

There has been some debate on whether these Page curves are relevant for realistic black holes where there is no bath. The following observations are uncontroversial. 
\begin{enumerate}
\item
Page \cite{Page:1993df} originally studied the algebra of all observables in the black-hole exterior. This can be defined precisely by starting with an HKLL-type \cite{Hamilton:2006az} representation of elementary field operators in terms of asymptotic operators and taking all possible products and linear combinations to obtain an algebra. The entropy of this algebra is always zero, if the black hole is in a pure state, and does not follow a Page curve in time. 
\item
It is possible to define other algebras whose entropy does obey a Page curve as a function of some parameter. Some examples are given in \cite{Laddha:2020kvp}. However, these algebras are somewhat contrived and the Page curve has no interpretation in terms of information ``emerging'' from the black hole.
\item
Some researchers believe that it might be possible to define a ``natural algebra'' whose entropy follows a Page curve that can be interpreted in terms of information emerging from the black hole. However, as of this writing, such an algebra has not been specified precisely.
\end{enumerate}

Returning to the case with a nongravitational bath, the discussion above helps to address a common criticism of these calculations \cite{Martinec:2022lsb}, which is that they do not explain how information enters the bath. Consider a two-step process. In the first step, one prepares a black hole in AdS. The principle of holography of information tells us that information about the state of the black hole is already present in observables near the boundary of the gravitational theory.  In the second step, one couples the system to a bath by turning on a coupling at the boundary of AdS.  This allows the information to flow locally from the boundary into the bath, and then from one part of the bath to the other. We note that information does not have to ``jump'' from the black-hole interior into the bath, and nor does this mechanism require any novel effects to mediate this process. 

\bibliographystyle{utphys}
\bibliography{references}
\end{document}